# SECULAR CHANGES IN SOLAR MAGNETIC FLUX AMPLIFICATION FACTOR AND PREDICTION OF SPACE WETHER

T.E.Girish and G.Gopkumar,
Department of Physics,
University College,
Thiruvanatnthapuram 695034
INDIA
E mail: tegirish5@yahoo.co.in

## ABSTRACT

We could infer a secular decreasing trend in the poloidal to toroidal solar magnetic flux amplification factor ($A_f$) using geomagnetic observations ( classic and IHV corrected aa indices) during the sunspot cycles 9-23. A similar decreasing trend is also observed for the solar equatorial rotation (W) which imply possibly a decrease in the efficiency of the solar dynamo during the above period. We could show correlated changes of Af and extreme space weather activity variations near earth since the middle of the 19[th] century. Indirect solar observations ( solar proton fluence estimates) suggests that the distinct enhancements in extreme space weather activity, $A_f$ and W found during sunspot cycles 10 to 15 is probably largest of that kind during the past 400 years. We find that the sunspot activity can reach an upper limit (R<300) when $A_f$ becomes unity. If the current sunspot cycle 24 turns out to be weak then very severe space weather conditions is most probable to occur during this cycle.

Key words: Flux amplification,solar dynamo, space weather, predictions,cycle 24

\

## 1. Introduction

The prediction of space weather conditions is intimately related to prediction and understanding the solar activity. Cyclic and long term changes in solar magnetic fields is known to affect several phenomena from sunspot activity changes to geomagnetic variations ( Lockwood et al,1999). According to solar dynamo models ( Choudhuri,2003: Schussler & Schmitt,2004) weak large scale poloidal magnetic fields during the sunspot minima is amplified by agencies like solar differential rotation to form strong toroidal magnetic fileds during sunspot maxima associated with active regions . Girish and Gopkumar( 2001) suggested that the magnitude of solar magnetic flux amplification changes from one sunspot cycle to another. They have also inferred secular changes in the poloidal to toroidal flux amplification factor during the sunspot cycles 9-23 using geomagnetic observations.

In this paper we have shown that the secular changes in extreme space weather conditions observed near earth during the past 150 years can be explained in terms of the secular changes in poloidal to toroidal solar magnetic flux amplification factor ($A_f$) inferred either from classic (Mayud,1980;Nevanlinna,2004) or IHV corrected geomagnetic aa indices ( Svalgaard and Cliver,2007). Linear relations are found between variations of solar magnetic flux amplification factor, solar equatorial rotation, sunspot activity and properties of large fluence ( >30MeV) solar proton events during the sunspot cycles 9-23. If these relations are valid for previous and future solar cycles then we show that we could make some interesting inferences. Sunspot number can reach an upper limit ( yearly mean R<300) when Af becomes unity. The variations of the fluence of >30MeV solar proton events inferred for the past 31 sunpot cycles by Shea etal ( 2008) from polar ice nitrate estimates suggest that the distinct enhancements in extreme space weather activity,Af and W found during the solar cycles 10-15 is unique and probably largest of that kind during the past 400 years. Finally if the current sunspot cycle turns out to be weak ( Svalgaard et al,2005:Choudhuri,2008) then occurrence of extreme space weather conditions ( great geomagnetic storms or large fluence solar particle events) is most probable during this cycle.

## 2. Secular decrease in solar magnetic flux amplification factor inferred from geomagnetic observations

Let us define the poloidal to toroidal solar magnetic flux amplification factor ( $A_f$) during a sunspot cycle ( Girish and Gopkumar,2001) as

$$A_f = B_t/B_p \qquad (1)$$

Here $B_p$ is the value of the solar poloidal magnetic flux density during a sunspot minima and $B_t$ is the maximum value of solar toroidal magnetic flux density during the following sunspot maxima.
Using geomagnetic aa indices , we can calculate $A_f$ using the relation

$$A_f = aa_{max}/aa_{min} \qquad (2)$$

Here $aa_{max}$ is the yearly mean aa index observed during sunspot minima proportional to $B_p$ and $aa_{max}$ is the yearly mean aa index observed during the sunspot maxima proportional to $B_t$

.Some examples of calculations of $A_f$ using classic aa indices ( available in NGDC website since the year 1868) are are given below:

$aa_{min}$ = 14  (1965) ,$aa_{max}$ = 22.5 ( 1968) and $A_f$= 1.6   for cycle 20

$aa_{min}$ = 20.2 ( 1977) ,$aa_{max}$ = 33.7 ( 1982) and $A_f$ = 1.67  for cycle 21.

The peaks in aa due to recurrent geomagnetic activity  as in 1974 will be avoided in our calculations  for selecting $aa_{max}$.

We can also calculate the flux amplification factor ( $A_{fc}$)  using IHV corrected aa indices ( available in Leif Svalgaard's website) for the years 1844- 2006. In Fig.1  we have plotted $aa_{min}$ , $aa_{max}$ and  values of $A_f$ calculated  using  classic as  well as  IHV corrected  aa indices  for the sunspot cycles 9-23. We have used yearly mean Helsinki aa indices  for the years 1844-1855 and for the year 1867. For sunspot cycle 10 ( 1856-1866) yearly aa indices are derived from yearly  Helsinki $A_k$ ( H) indices using the following  regression relation ( Nevanlinna,2004)

aa  =   1.31 $A_k$ ( H) - 0.2        (3)

The   IHV corrected $aa_{min}$  and $aa_{max}$ are found by adding 6nT to the corresponding   classic aa values  for the  cycle 10.

The parameters $aa_{min}$  and $aa_{max}$ ( classic and corrected)  in general show a secular increasing trend  ( Fig 1a and 1b) while the flux amplification parameters  $A_f$ ($A_{fc}$)  show  a secular decreasing trend ( Fig 1c). We can also find a high correlation between variations of  $A_f$  and $A_{fc}$ ( r = 0.9).  Distinct peaks are found in   $A_f$ during the sunspot cycles 10 and 12. The solar magnetic flux amplification factor inferred from classic indices ( Af)  is reduced to  half  between the sunspot cycles 12 to 23. . But the Afc calculated from corrected aa indices show a decrease of only 40%  during the above period

### 3. Relations of flux amplification factor with sunspot activity and solar rotation changes during cycles 9-23

In Fig 2 (a)  we have plotted the maximum yearly mean sunspot number ( international sunspot numbers available in NGDC website)  $R_{max}$ along with  $A_f$  for the solar cycles 9-23. $R_{max}$ show a steady increasing trend between sunspot cycles 14-19. We can also find statistically significant anti correlation between inferred flux amplification factor ($A_f$ or $A_{fc}$)  variations  and $R_{max}$ changes  during cycles 9-23 ( see Table 1). The following regression relations are  found between these parameters.

$A_f$ =3.1389 -  0.0095 $R_{max}$        (4)

$A_{fc}$ =2.2219 - 0.0045 $R_{max}$     (5)

In Fig 2(a) we have plotted solar equatorial rotation  rate ( W)  inferred from sunspot observations ( Javaraiah,2003) during the sunspot cycles 12-23. We can notice a secular decreasing trend in W similar  to $A_f$ during this period. For comparison we have plotted time of rise ( $T_r$ in months) of sunspot activity during the sunspot cycles 9-23 (Kane,2008). It is interesting to observe that both W and $T_r$ reached a well defined minima during the sunspot cycle 22. If  we  apply the condition $A_f$ =1 in the regression relations  (2) or (3) we can find an upper limit to Rmax  or sunspot activity . The maximum value of $R_{max}$ derived from relation (2) is found to be 225 and from relation (3) is 272. Such a situation arises during grand solar maximum epochs known from radio carbon studies of past solar activity.

**4. On the physical basis of the secular changes in extreme space weather activity near earth**

Great geomagnetic storms and large fluence solar proton events ( SPE) are two important manifestations of extreme space weather activity ( Cliver and Svaalgard,2004) observed near earth. McCracken etal ( 2001) inferred large fluence solar proton events ( energy > 30 MeV , fluence > 2 X $10^9$ particles/$cm^2$ ) using nitrate estimates from polar ice cores during the years 1561-1950 A.D . The number of such large fluence SPE's per sunspot cycle ( $N_s$) is plotted in Fig.2 along with Af for the sunspot cycle 9-23. The SPE data during 1951-2004 is adopted from Shea etal ( 2006). It is interesting to observe a maxima in $N_s$ during the cycle 13 and a decreasing trend between cycles 14-22 similar to $A_f$. The correlation coefficient between $A_f$ and $N_s$ variations is found to be statistically significant.But the correlation of Ns with solar equatorial rotation( W) changes is still higher ( Table 1) . The following regression relations are obtained.

$N_s$ = 1.4961 $A_f$ - 0.6828      (6)

W  =  0.004  Ns + 2.9195      (7)

During cycles 10-15 with an exception of cycle 11 we can find Af >2 ( Afc >1.75) and during cycles 16-23 we find Af < 2 ( Afc < 1.75) Secular changes is also found in the occurrence of great geomagnetic storms observed during this period. The intensity of a geomagnetic storm is best measured using low latitude Dst index. Conidering Dst data available since the year 1932 it is found that the March 14 1989 storm has the maximum intensity of 589 nT ( Cliver and Svaalgard,2004).
 Approximate values of intensity ( Dst) of geomagnetic storms prior to the year 1932 can be inferred from H range of low-latitude stations like Bombay in India or from the observation of lowest latitude of aurora ( Silverman,2006) during such intense storms. In Table 1 we have given a list of great geomagnetic storms occurred during sunspot cycles 10-15 with inferred intensity significantly greater than that of the March 1989 storm. Thus the secular decrease in inferred solar flux amplification factor can provide a physical basis for the observed decrease in the occurrence of extreme space weather events observed near earth during the recent sunspot cycles (16-23) compared to the sunspot cycles 10-15.

**5.. Space weather predictions during cycle 24**

The forecast of maximum yearly mean sunspot number ( Rmax) ranges from 67-150 for the current sunspot cycle 24 ( Svalgaard et al,2005: Dikpati and Gilman,2008: Choudhuri,2008;Brajsa et al,2009). Statistically significant correlations between different parameters given in Table 1 will be helpful for space weather predictions. Making use of linear regression relations between (a) Af & Rmax ( equation 1) and (b) Af & Ns ( equation 3) we have estimated probable values of Af and Ns expected for the sunspot cycle 24 depending on the maximum value of sunspot number which is given in Table 2. From the total fluence (Fp) data of >30 MeV solar proton events for the
sunspot cycles 9-23 ( Shea etal,2008) we can find the following regression relation between the number of large fluence SPE's ( $N_s$) and $F_p$ as

$F_p$  =  0.5313 $N_s$ + 0.2336           (8)

Making use of the relation (6) we have also calculated the expected fluence Fp of SPE's
corresponding to different Ns values inferred for cycle 24 and given in Table 2. We can find that if $R_{max}$ < 100 as predicted by Svaalgard and others then $A_f$>2 and the probability of occurrence of extreme space weather activity will be high during this cycle. In contrast with this if $R_{max}$>150 as predicted by Dikpati and Gilman, then the space weather conditions during the cycle 24 will not be severe than the previous cycle 23.

**6. Discussion**

The inferred secular decrease in solar magnetic flux amplification factor (inferred from classic as well as corrected aa indices) during cycles 9-23 can be related to a possible secular decrease in the efficiency of the solar dynamo. Thus secular decrease in extreme space weather activity during the recent fifty years inspite of the increasing trend in solar activity can be related to $A_f$ decreases.
When $A_f > 2$ during cycles 10-15 we could find at least five solar proton events ( > 30 MeV ) exceeding the fluence of the August 1972 event ( Shea at al,2006). In fact the solar proton event associated with the Carrington solar flare during September 1859 has an inferred fluence of three times that of the modern August 1972 event. At least six great geomagnetic storms occurred during sunspot cyles 10-15 with an inferred intensity much greater than the March 14 1989 geomagnetic storm. Apart from the five listed in Table 1 . the February1872 geomagnetic storm can be included for which aurora is observed close to magnetic equator ( Silverman,2008) with an inferred Dst of 1020 nT ( Lakhina et al,2005).

Using linear regression relations ( 1,3 and 4) we could predict the values of $A_f$ and the probable number ( $N_s$) and fluence ( $F_p$) of > 30 MeV solar proton events for different values of maximum yearly sunspot number fpr a given sunpot cycle. If solar cycle 24 turns out to be a weak sunspot cycle with Rmax around 75 as predicted by Svaalgard et al ( 2005) then there is high probability of observation of extreme space weather activity in this cycle. Solar proton events (>30 MeV) with fluence exceeding the August 1972 event and geomagnetic storms with intensity significantly greater than that of March 1989 storm can be then expected to occur during this cycle.
The observed anti-correlation between flux amplification factor and maximum yearly sunspot number discussed in section 3 can be a basic property of the solar dynamo. The inverse relation between Rmax and time of rise of sunspot activity ( Tr) is well known. Using the relevant data for cycles 9-23 we can find the following regression relation

$R_{max}$ = 294.73 -3.7121 $T_r$ (9)

If we apply the condition $T_r$=0 in equation (6) then we can find an upper limit for $R_{max}$ as 295.
The upper limits of Rmax estimated from linear relations with Af is found to be 231 and that with Afc is 272 . Thus it is reasonable to assume that the upper limit of yearly mean sunspot number is less than 300. For grand solar activity maximum epochs such as the medieval maximum observed between 1100 -1200 AD , the conditions $T_r$ =0 and $A_f$=1 are both seem to be applicable. However the sunspot numbers estimated for the medieval maximum from inferred cosmic ray intensity near earth by Usoskin etal (2003) is too low compared to our estimates of the upper limit!

Several authors have suggested the need for correcting calibration errors in the classic aa indices since it has probably over estimated long term ( centennial scale) increases in geomagnetic activity . Since geomagnetic records archive valuable information about the sun and solar activity in the past , the corrections in them should be done with utmost care so that changes within a sunspot cycle are not affected. Geomagnetic observations from Indian observatories in Trivandrum ( near magnetic equator) and Bombay ( low latitude) suggest that (Eapen,2009) that the sunpot cycle 10 ( 1856-1866) is a cycle with exceptional geomagnetic activity with occurrence of at least five geomagnetic stroms exceeding the intensity of modern March 14,1989 storm, of which three happened during the year 1859. In this context we have found that Helsinki Ak(D) indices underestimated the sunspot cycle change in geomagnetic activity during the cycle 10 and Helsinki Ak(H) indices ( Nevanlinna,2004) is a better substitute for calculation of $A_f$ during cycle 10. From Table 1 we can see that $A_f$ calculated from classic aa indices show a better correlation with sunspot activity related parameters ( $R_{max}$,$T_r$ and W) and extreme space weather activity parameters ( $N_s$ and $F_p$) compared to $A_{fc}$ calculated from IHV corrected aa indices. The sunspot cycle change of geomagnetic activity ( measured as a ratio in Af) , happening over a period of within 7 years is not likely to be affected by real calibration errors if any in classic aa indices in the past. Thus Af calculations using classic aa indices can be still relevant.

In Fig 3 (a)we have plotted total fluence of ( >30 MeV) of solar proton events ( Fp) in a sunspot cycle inferred from polar nitrate estimates by Shea etal ( 2008) since the beginning of sunpot observations ( 1610-2007 AD). The data covers 31 sunspot cycles and can be considered as longest homogeneous record of space weather activity observed near earth. Using linear relations found between Ns,Fp,Af and W during sunspot cycles 9-23 it is possible to estimate flux amplification factor and solar equatorial rotation rate back in time ( since sunspot cycle -12 starting in 1610) assuming that the relevant correlations are valid through this long period. Such an estimate is shown in Fig 3(b).The results suggest that sunspot cycles 10 to 15 is a period of unusual increase in space weather activity caused probably by an extraordinary increase in solar magnetic field flux amplification factor ( Af) and solar equatorial rotation rate ( W) during the past 400 years

**7.Conclusions**

(i) We could find secular decrease in poloidal to toroidal solar magnetic flux amplification factor (Af) inferred from geomagnetic observations during the cycles 9-23 .This suggests that the efficiency of the solar dynamo is probably decreasing since the sunpot cycle 12. A decreasing trend found for solar equatorial rotation rates during the cycles 12-23 provide an independent support to this result.

(ii) Secular changes in solar dynamo is also reflected in the heliosphere.We could find a secular decrease in extreme space weather activity observed near earth since the middle of the 19$^{th}$ century correlated with Af changes. Indirect solar observations ( solar proton fluence estimates) suggests that the distinct enhancements in extreme space weather activity , Af and W found during sunspot cycles 10-15 is probably largest of that kind during the past 400 years.

(iii) Linear relations between solar magnetic flux amplification factor, sunspot activity and properties of large fluence solar proton events found during solar cycles 9- 23 suggest that if the current sunspot cycle 24 turns out to be a weak sunspot cycle (Rmax < 100 ) then the observation of extreme space weather conditions is most probable in this cycle.

( iv) Inverse relations found between sunspot activity and Af can be used to determine an upper limit for maximum yearly sunspot number ( R<300).

**Aknowledgements**

The authors wish to thank the International Astronomical Union for providing financial assistance to attend the IAU Symposium 257: *Universal Heliophysical Processes* held in Greece during September 2008 where a preliminary version of this paper has been presented. The authors are grateful to
NGDC,Boulder,USA for maintaining solar and geomagnetic data. We also thank Prof.C.Radhakrishnan Nair for useful discussions.

**Figure Captions**

Fig 1. Variations of ( a) $aa_{min}$ (b) $aa_{max}$ (c) Flux amplification factors derived from classic and IHV corrected geomagnetic aa indices during the sunspot cycles 9-23

Fig 2. Variations of (a) Maximum sunspot number ( $R_{max}$) and flux amplification factor ($A_f$)   (b) Time of rise of sunspot activity ( $T_r$) and solar equatorial rotation rate during the sunspot cycles 9-23.

Fig 3.   Variations of number of large fluence ( >30 MeV) solar proton events per cycle  ( $N_s$) and flux amplification factor ( $A_f$) during the sunspot cycles 9-23.

Fig 4. (a) Variations of the  total fluence ($F_p$)  of  large ( >30 MeV) solar proton events inferred by Shea et al ( 2008) from nitrate estimates from polar ice cores during the sunspot cycles -12 ( starting from 1610)  to 23 ( ending on 2007).
         (b) Variations of inferred values of solar magnetic flux amplification factor (Af)  and solar equatorial rotation rate ( W) during the sunspot cycles -12 to 21. Values Af and W  during cycles -12 to 8 are inferred from solar proton fluence data above . Values of Af  are inferred from  from geomagnetic data and W from sunspot observations during  the sunspot cycles 9-21.

**Table 1. Correlation of flux amplification parameters and solar equatorial rotation with different solar terrestrial phenemena. Statstically significant correlations are given in bold figures**

| Parameter | $R_{max}$ | $T_r$ | W | $N_s$ | $F_p$ |
|---|---|---|---|---|---|
| $A_f$ | **-0.61** | **0.65** | **0.61** | **0.52** | 0.39 |
| $A_{fc}$ | **-0.54** | 0.46 | 0.53 | 0.43 | 0.28 |
| W | **-0.64** | 0.55 | | **0.71** | 0.50 |

**Table 2. Selected great geomagnetic storms during sunspot cycles 10-15 with intensity significantly higher than the March 1989 storm**

| Date of storm | Sunspot cycle ($R_{max}$) | Inferred Dst | Reference | Af during the cycle |
|---|---|---|---|---|
| 1859 Sep 2 | 10 (95.8) | -1700 nT | Tsurutani etal ( 2003) | 3 |
| 1859 Oct 12 | 10 (95.8) | -980 nT | Lakhina et al ( 2005) | 3 |
| 1903 Oct 31 | 14 (63.5) | -820 nT | Lakhina et al ( 2005) | 2.48 |
| 1909 Sep 25 | 14 (63.5) | -1000 nT | Silverman ( 1995,2006) | 2.48 |
| 1921 May 13 | 15 (103.9) | -900 nT | Kappenman ( 2006) | 2.6 |

**Table 3. Predicted values of flux amplification factor ( Af), number of large fluence ( >30 MeV) solar proton events ( Ns) and its corresponding fluence ( Fp) for different possible values of maximum sunspot number during sunspot cycle 24.**

| $R_{max}$ | $A_f$ | $N_s$ | $F_p$ ($10^{10}$ particles cm$^{-2}$) |
|---|---|---|---|
| 60 | 2.57 | 3.12 | 1.89 |
| 70 | 2.47 | 2.98 | 1.82 |
| 80 | 2.38 | 2.84 | 1.74 |
| 90 | 2.28 | 2.70 | 1.67 |
| 100 | 2.18 | 2.57 | 1.59 |
| 110 | 2.09 | 2.43 | 1.52 |
| 120 | 1.99 | 2.29 | 1.45 |
| 130 | 1.90 | 2.15 | 1.38 |
| 140 | 1.80 | 2.01 | 1.30 |
| 150 | 1.71 | 1.87 | 1.23 |

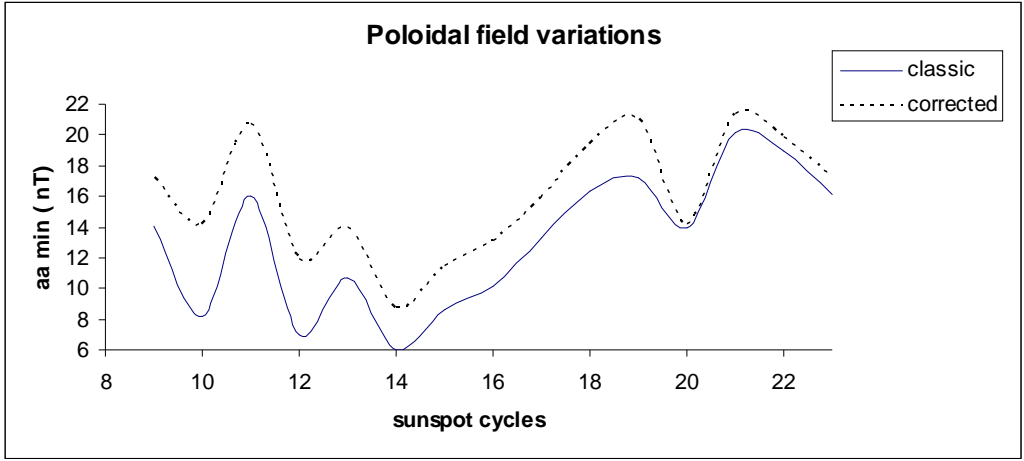

**(a)**

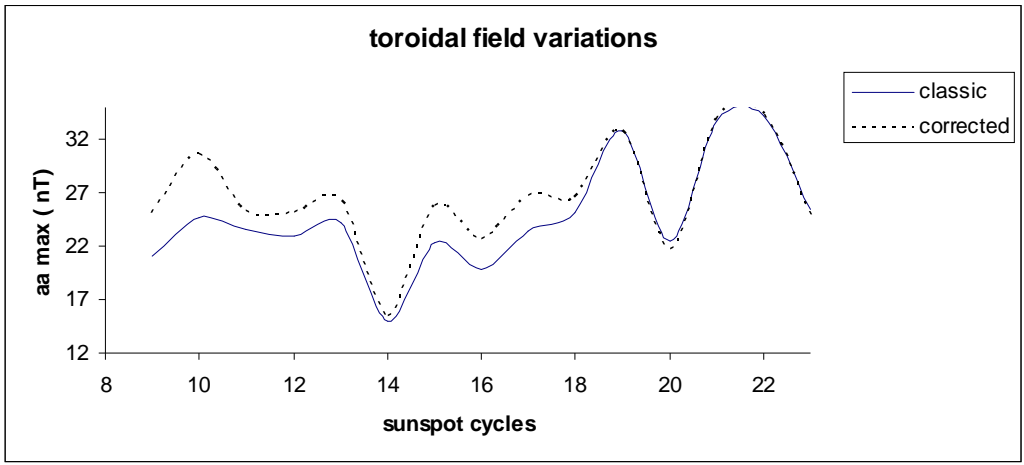

(b)

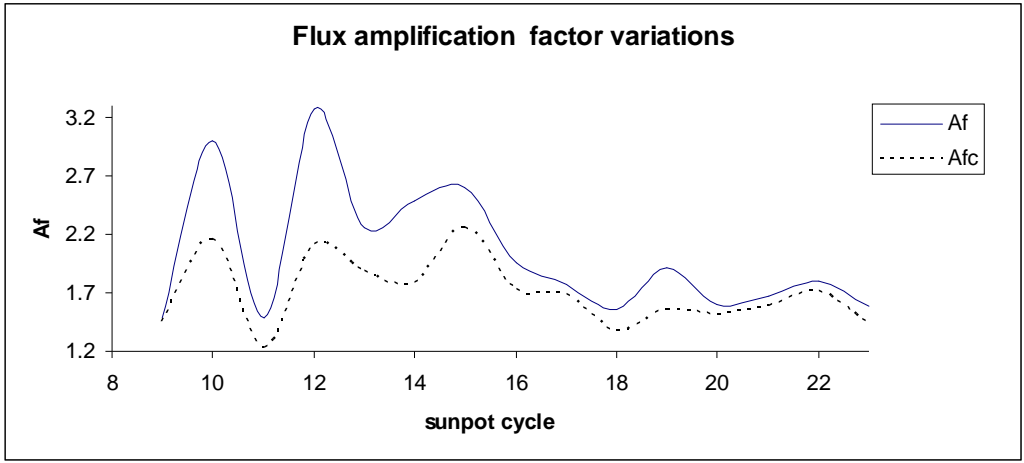

(c)
**Fig 1**

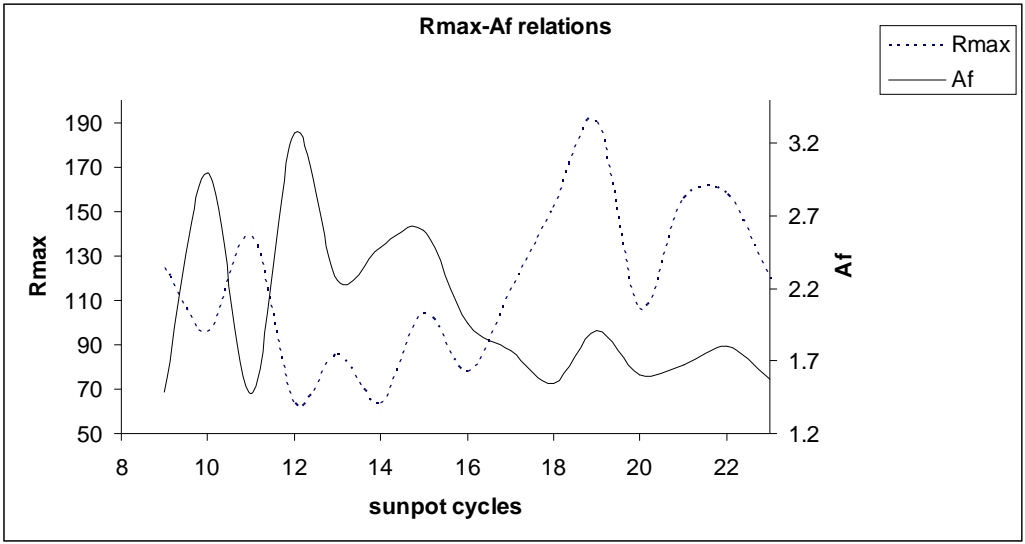

**(a)**

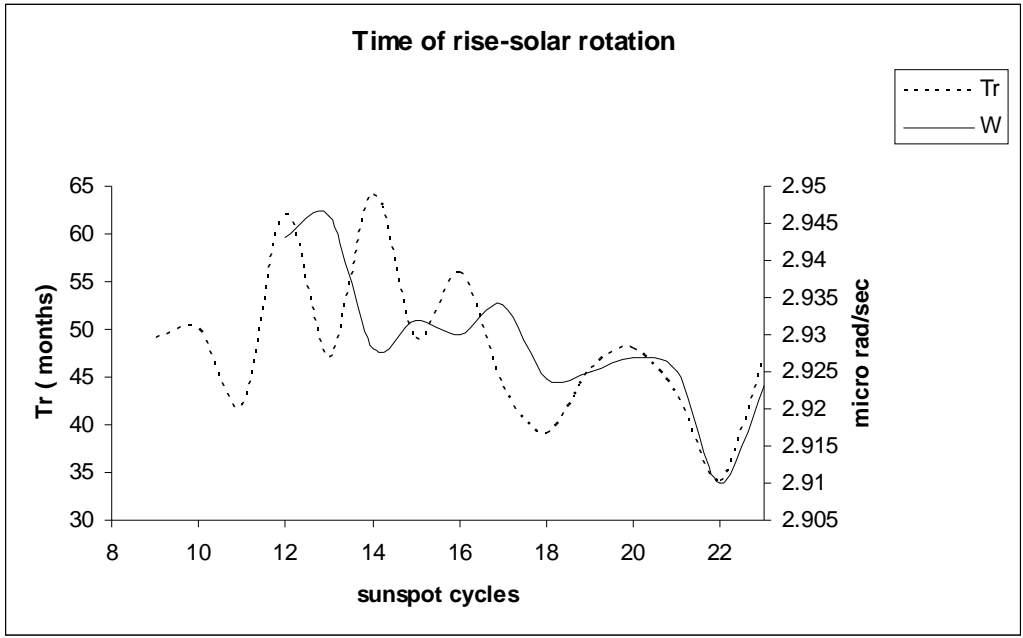

**(b)**

**Fig 2**

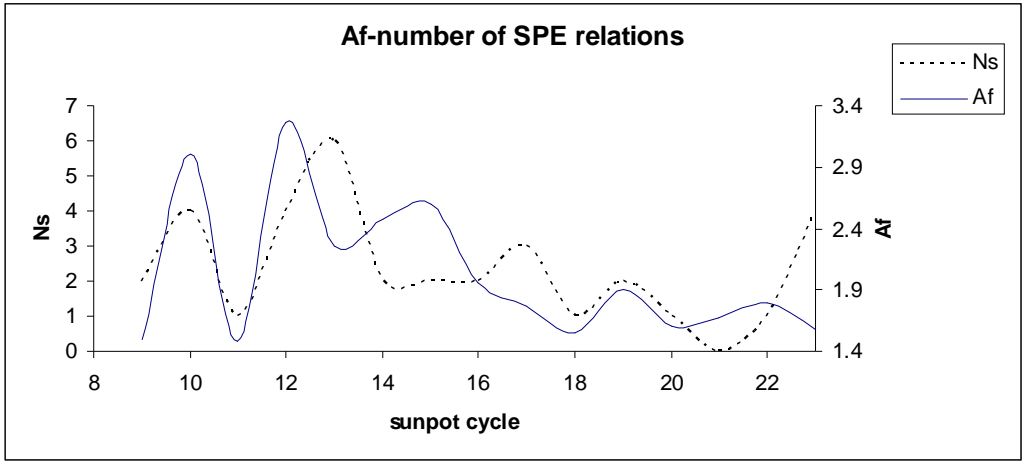

**Fig 3**

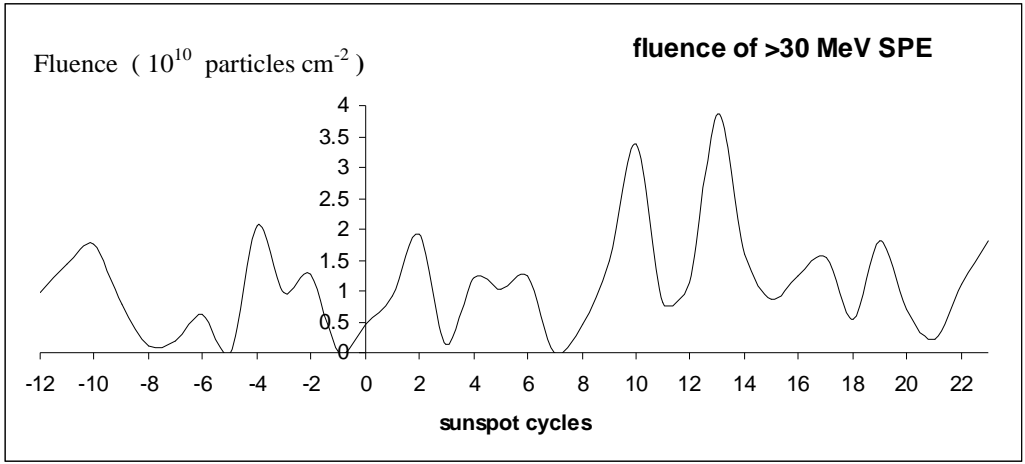

(a)

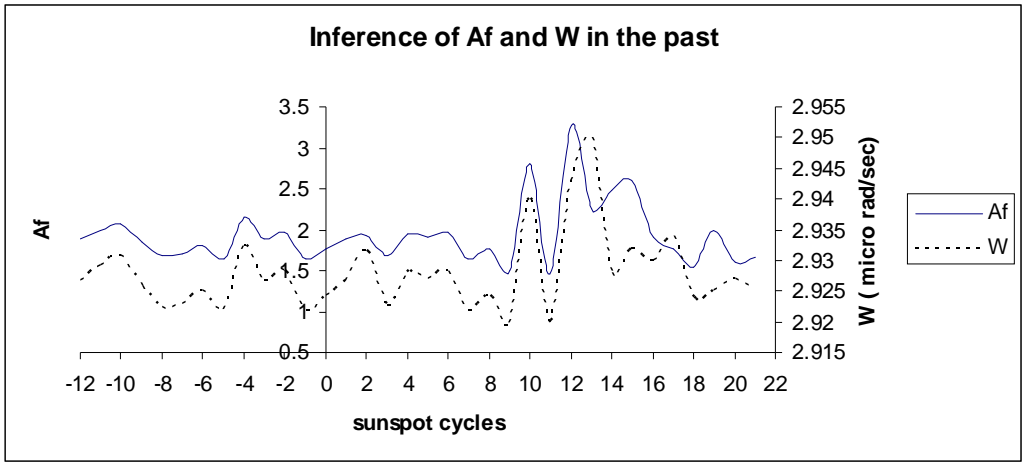

(b)

**Fig 4**